\documentclass[conference,a4paper]{IEEEtran}

\usepackage{xcolor}
\usepackage{balance}
\usepackage{soul}
\usepackage{amsmath}
\usepackage{algpseudocode}
\usepackage{algorithm}
\usepackage{amssymb}

\ifCLASSINFOpdf
  \usepackage[pdftex]{graphicx}
\else
\fi

\ifCLASSOPTIONcompsoc
 \usepackage[caption=false,font=normalsize,labelfont=sf,textfont=sf]{subfig}
\else
 \usepackage[caption=false,font=footnotesize]{subfig}
\fi

\begin{document}
\title{A Novel Blind Adaptive Beamformer with Robustness against Mutual Coupling and Miscalibration Effects}

\author{\IEEEauthorblockN{
M. Yaser {YAĞAN}\IEEEauthorrefmark{1}\IEEEauthorrefmark{2},   
Ahmet F. {COŞKUN}\IEEEauthorrefmark{1},
Ali E. {PUSANE}\IEEEauthorrefmark{2}    
}                                     
\IEEEauthorblockA{\IEEEauthorrefmark{1}
H\.{I}SAR Lab. @Informatics and Information Security Research Center (B{\.{I}}LGEM), T{\"{U}}B{\.{I}}TAK, Kocaeli, Turkey}
\IEEEauthorblockA{\IEEEauthorrefmark{2}
Department of Electrical and Electronics Engineering, {Boğaziçi} University, İstanbul, Türkiye}
\IEEEauthorblockA{ \emph{yaser.yagan@tubitak.gov.tr/yaser.yagan@boun.edu.tr, ahmet.coskun@tubitak.gov.tr, ali.pusane@boun.edu.tr} }
}

\maketitle

\begin{abstract}
Beamforming techniques utilized either at the transmitter or the receiver terminals have achieved superior quality-of-service performances from both the multi-antenna wireless communications systems, communications intelligence and radar target detection perspectives. Despite the overwhelming advantages in ideal operating conditions, beamforming approaches have been shown to face substantial performance degradations due to unknown mutual coupling effects and miscalibrated array elements. As a promising solution, blind beamformers have been proposed as a class of receiver beamformers that do not require a reference signal to operate. In this paper, a novel gradient-based blind beamformer is introduced with the aim of mitigating the deteriorating effects of unknown mutual coupling or miscalibration effects. The proposed approach is shown to find the optimal weights in different antenna array configurations in the presence of several unknown imperfections (e.g., mutual coupling effects, miscalibration effects due to gain and phase variations, inaccurate antenna positions). By providing numerical results related to the proposed algorithm for different array configurations, and bench-marking with the other existing approaches, the proposed scheme has been shown to achieve superior performance in many aspects. Additionally, a measurement-based analysis has been included with validation purposes.
\end{abstract}

\vskip0.5\baselineskip
\begin{IEEEkeywords}
 antenna array processing, receive beamforming, mutual coupling, array calibration imperfections.
\end{IEEEkeywords}

\section{Introduction}
Beamforming is a fundamental method that has been considered in wireless communication research communities for decades. The purpose of different beamforming techniques is to exploit antenna arrays to achieve reliable links with low power consumption. Consequently, beamformers can be realized at transmitters, receivers, or both transmitters and receivers in analog, digital, or hybrid forms. At the transmitter side, complex weights are calculated and multiplied by the signal fed to the array such that the propagated electromagnetic waves add up constructively in the desired beam direction and destructively in other (null) directions. On the other hand, a similar formulation can be implemented at the receiver to maximize the receiver array's gain in a specific direction and force it to approach zero in other interference directions. {A high-performance} beamformer {strictly} requires perfect channel estimation at the receiver and undelayed feedback to the transmitter such that the complex weights {could} be jointly optimized.

Receiver beamforming algorithms {might} be classified in three forms:According to training necessity, there are data-aided and blind beamformers,for weight calculation, deterministic and iterative algorithms exist,and according to flexibility, there are fixed and adaptive beamformers.
While adaptiveness is a certain requirement in mobile communications, iterative and deterministic algorithms can be compared in terms of computational complexity and its trade-off with the convergence of the iterative algorithm. Furthermore, a well-performing blind beamformer is preferable over a data-aided beamformer, since it will significantly reduce the {signaling} overhead.
A data-aided beamformer calculates the weights numerically by optimizing a divergence measurement between the received signal and a reference signal {via a mean square error (MSE) evaluation approach}. However, blind beamformers {search for the} weights that maximize specific criteria, {such as} the signal-to-noise ratio (SNR) or signal-to-interference plus noise ratio (SINR). This is usually achieved by estimating the angle of arrival (AoA) of the desired signal and interference, then calculating the weights that give high gain at the desired direction and form nulls in the other directions. These methods suffer from many drawbacks arising from the fact that a precise AoA estimation is required and this depends essentially on the array factor (AF). Consequently, unknown mutual coupling (MC) or miscalibrated array (MA) limits the performance of such beamformers significantly.

{Few studies have focused on the problem of blind beamforming in the presence of unknown MC}. In \cite{ULAmc}, a beamforming algorithm {has been} developed for ULAs with unknown MC depending on the structure of the mutual coupling matrix (MCM). {The numerical results have shown that the proposed algorithm achieves a better performance in the presence of unknown MC when compared to other approaches.} {Within the same context, the recent work in \cite{3Dmc}} has proposed an approach to improve the estimation of the steering vector. The approach was applied to conventional beamformers and provided performance improvement.

Aside from these methods, a well-known algorithm for blind beamforming, previously used for blind equalization \cite{CMAequalizer}, is the constant modulus algorithm (CMA) \cite{CMA2}. The CMA basically aims to numerically maximize the constant modulus property of the desired {signals, whereas its solution is limited to only constant-amplitude signals (e.g., phase- and frequency-modulated ones). Additionally, other major deficiencies} of this algorithm {are shown to be} the high computational complexity and slow convergence \cite{CMAdif}. As a result, {in order to address its basic drawbacks,} CMA has been studied extensively and {that has yielded to several modified implementations with the aim of providing further enhancements}. For example, the authors in \cite{CMAdif} proposed combining the CMA with a data-aided scheme to exploit the advantages of both approaches. In \cite{CMA5}, a modified CMA with analog beamforming {has been} proposed, and {considerable enhancements in the overall bit error rate performance of a communications scheme have been} achieved. Other works \cite{CMA3,CMA4} focusing on the mathematical structure of the problem and reducing its complexity {have shown} significant reductions in {computational complexity or the required} processing power {by performing an extensive comparison on several stochastic gradient-based algorithms.} {\color{black}{A recent work \cite{iceict} has proposed another approach that aims to maximize the constant modulus criteria in a deterministic manner. Despite the substantial achievements with respect to other schemes in comparison, the complexity of this approach could easily be prominent as a bottleneck, since it includes eigenvalue decomposition and matrix inversion operations. In order to speed up the convergence phase, the authors in \cite{Hanning} have proposed employing a Hanning window on the updated weights. However, using Hanning window is shown to limit the degrees of freedom and results in an degradation offset when compared to the optimum solution.}}

In this paper, a blind iteratively adaptive beamformer for a single source scenario is proposed. The proposed beamformer is gradient-based and not limited to {signal subspaces oriented by an inflexible AF}. {Since the proposed scheme has no dependencies on the array geometry, it is expected to be much more robust against unknown MC and MA impairments.} Simulation results have shown that the proposed {approach} maximizes the received signal's power {for} different {array configurations in the presence of unknown antenna array imperfections (AAI)}. {By providing an extensive study on the performance evaluation and comparison to other approaches together with measurement-based validation process, the proposed algorithm has been shown to achieve the optimal solution with reduced complexity and number of iterations. Besides, thanks to its adaptive structure, the algorithm is able to effectively update the combining weights due the AoA variations while facing no performance degradations.} 

The {remainder} of this paper is organized as follows: The system model and mathematical formulation are explained in Section II, the proposed {algorithm} is presented in Section III, numerical results and comparisons are given in Section IV, and Section V concludes the paper.

\section{System Model}
\begin{figure}
\centering
\includegraphics[width=0.7\columnwidth]{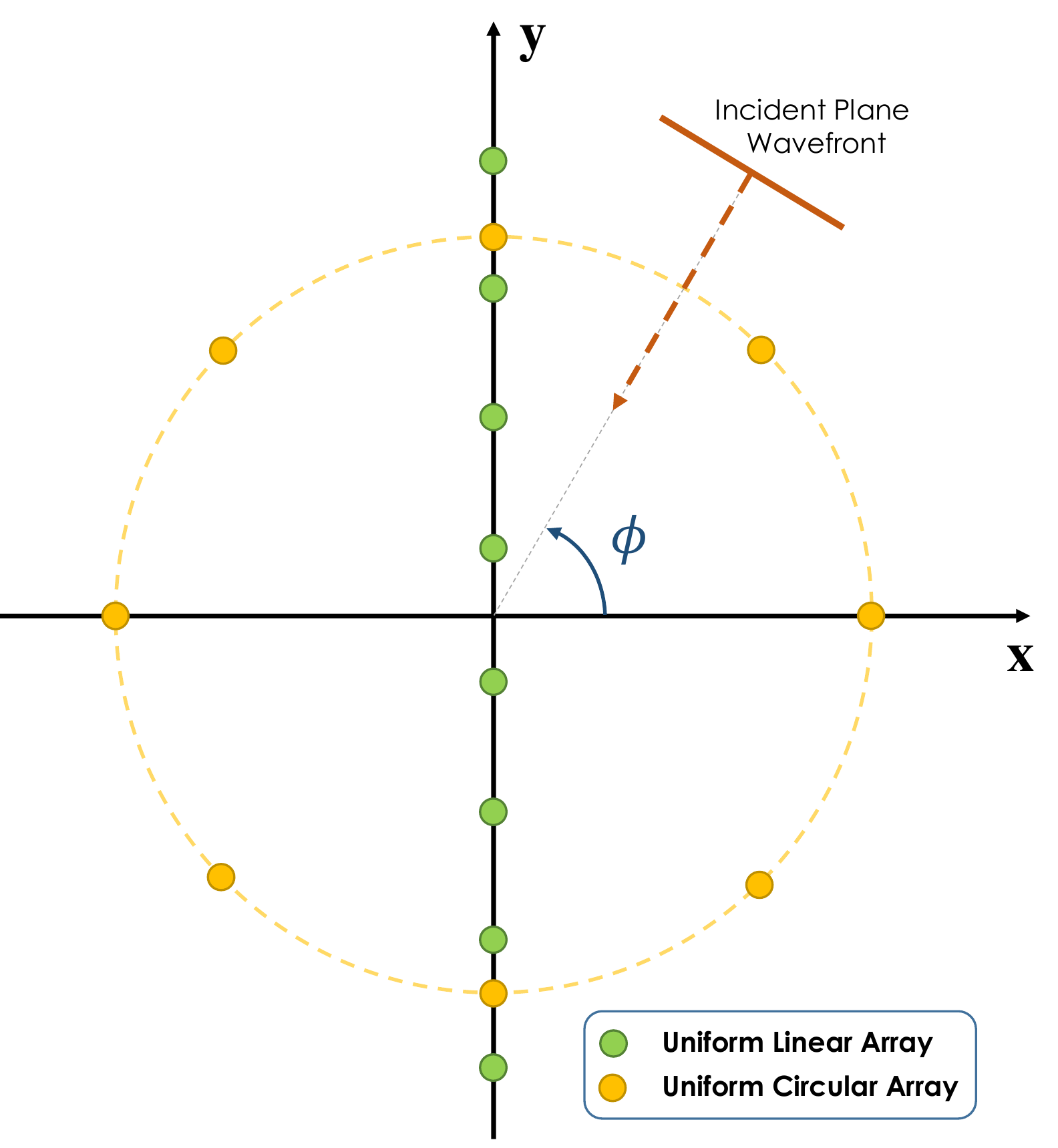}
\caption{Eight-element uniformly spaced linear and circular arrays receiving signals from a single emitter at
direction $\phi$.}
\label{fig:AoA}
\end{figure}

{This paper focuses on} an arbitrary planar antenna array composed of $M$ elements ({as exemplified in Fig. \ref{fig:AoA}} for uniform linear array (ULA) and uniform circular array (UCA) cases) {that could be} described by the element positions in {two-dimensional Cartesian} coordinate system as $(p_{ix},p_{iy}), i=1,2,...,M$. Accordingly, for an incident {planar} wave, the azimuth AoA $\phi$ can be defined as the angle between the plane's {normal vector} and the {$\bf{x}$} axis, as shown in Fig. \ref{fig:AoA}.
{The transmitted radio-frequency (RF) signal will be sensed at each antenna element with systematically varying time lags}. As a function of the incidence signal direction $\phi$, the time lags of the received signals {$\tau_i(\phi)$ could be easily expressed in terms of the elements' positions $(p_{ix},p_{iy})$ relatively to the phase center of the array{\footnote{The phase center might be selected arbitrarily with no restrictipn, but for the sake of simplicity, the origin of the Cartesian coordinate system is selected.}.}} Hence, under noise-free and lossless transmission conditions considered to describe the signal model, the received signal vector consisting of the time-domain complex RF signal replicas received by all elements might be expressed as 
\begin{equation}
    \mathbf{r}(t;\phi) = \begin{pmatrix}
    s(t-\tau_1)e^{j2\pi f_c(t-\tau_1)} \\
    s(t-\tau_2)e^{j2\pi f_c(t-\tau_2)}\\
    \vdots \\
    s(t-\tau_M)e^{j2\pi f_c(t-\tau_M)}\\
    \end{pmatrix},
\end{equation}
{where $s(t)$ is a baseband signal occupying a bandwidth of $B$, and $f_c$ denotes the RF carrier frequency. As long as the narrowband signal assumption (i.e., $B<<f_c$) holds true, each signal replica might be approximated as:
$s(t-\tau_M)e^{j2\pi f_c(t-\tau_M)}\approx s(t)e^{j2\pi f_c(t-\tau_M)}$. Hence the received signal vector in (1) might be rewritten as
\begin{equation}
    \mathbf{r}(t;\phi) =s(t)e^{j2\pi f_c t} \begin{pmatrix}
    e^{-j2\pi f_c\tau_1} \\
    e^{-j2\pi f_c\tau_2}\\
    \vdots \\
    e^{-j2\pi f_c\tau_M}\\
    \end{pmatrix}\triangleq \underbrace{s(t)e^{j2\pi f_c t}}_{x(t)}{{\bf{a}}(\phi)}.
\end{equation}
In (2), $x(t)$ is the time-domain complex RF signal and ${{\bf{a}}(\phi)}$ is the array manifold (or the array factor, AF) that could fully characterize the overall response of the antenna array at a specific frequency and towards an incidence signal direction $\phi$. Since the RF signal part is common for all array elements, carrying out the investigation as focused on the baseband equivalents would be beneficial for tractability.} Accordingly, the digital baseband signal vector received by the array is given as
\begin{equation}
    {\mathbf{r}[n]} = s[n] \mathbf{a}(\phi) + \mathbf{\eta}[n],
\end{equation}
where $s[n]$ is the transmitted signal symbol at time instance $n$ and $\mathbf{\eta}$ is an $M\times 1$ vector of uncorrelated {zero-mean} additive white Gaussian noise {samples} with {a variance of $\sigma^2$}. 

{In the presence of antenna array impairments such as unknown MC, MA, and inaccurate positioning of antenna elements, the array manifold would face additional phase terms in its each element.} This would correspond to an array manifold that considers imperfections and is given as
\begin{equation}
    {\mathbf{a}_{I}(\phi)} = \begin{pmatrix}
    e^{j2\pi f_c \tau_1(\phi) + j\alpha_1} \\
    e^{j2\pi f_c \tau_2(\phi) + j\alpha_2}\\
    \vdots \\
    e^{j2\pi f_c \tau_M(\phi) + j\alpha_M}\\
    \end{pmatrix},
\end{equation}
{where $\alpha_i$, $i\in{1,2,\dots,M}$, denote the additional deteriorating phase terms caused by the impairments that would be simply equal to zero in ideal conditions ($\alpha_i=0$, $\forall i$), yielding $\mathbf{a}_{I}(\phi)\equiv \mathbf{a}(\phi)$.}

The beamformer multiplies the received signal vector {$\mathbf{r}[n]$} by a vector of complex weights $\boldsymbol{\omega}^H$ to give the output
\begin{equation}
    y[n] = \boldsymbol{\omega}^H\mathbf{r}[n]=\sum_{i=1}^{M}{{\omega}_i^* r_i[n]},
    \label{eq:yn}
\end{equation}
such that the output power is maximized. Hence, the optimum {weight vector $\hat{w}_{opt}$} is found by solving the optimization problem
{
\begin{flalign}
    \hat{\boldsymbol{\omega}}_{opt}\triangleq  &\arg\max_{\boldsymbol{\omega}}{\left\|\boldsymbol{\omega}^H\mathbf{r}[n]\right\|}^2\nonumber\\
    =&\arg\max_{\boldsymbol{\omega}}{\left\|\boldsymbol{\omega}^H(s[n] \mathbf{a}_{I}(\phi) + \mathbf{\eta}[n])\right\|}^2
\end{flalign}
}

With the constraint $\|\boldsymbol{\omega}\|=1$, the solution to {the optimization} problem {would easily be obtained as} 
{
\begin{equation}
    \hat{\boldsymbol{\omega}}_{opt} = \mathbf{a}_{I}(\phi).
    \label{eq:optimum_solution}
\end{equation}
}
{Here, it is clearly seen that,} given the array geometry and phase imperfection values $\alpha_i$, $\forall i$, {the overall problem} reduces to AoA estimation. Furthermore, for a blind beamformer, estimating the AoA with unknown {AAI} limits {the end-to-end} performance. Conventional AoA estimation methods, {such as} MUSIC, {search} for {the} solution in the subspace of $\mathbf{a}(\phi)$, thus they theoretically fail to solve this problem, since {$\mathbf{a}_{I}(\phi)$} lies outside. However, a numerical method like gradient descent {will not be} constrained within that subspace and {could still} search the {entire} $M$-dimensional space to reach the optimal solution. 

In the following section, {the proposed gradient-based power maximizing algorithm that aims to find the optimal weights for the blind beamformer is introduced.}. The objective function is the direct estimation of the received signal power and the gradient with respect to beamformer weights is calculated at each iteration to update the weights.
\section{Gradient-based Blind Adaptive Beamformer}
\subsection{Gradient Derivation}
The average power of the received signal can be estimated as
\begin{equation}
    \hat{P} = \frac{1}{N} \sum_{n=0}^{N-1}{|y[n]|}^{2} = \frac{1}{N} \sum_{n=0}^{N-1} y[n]{{y^*}}[n],
    \label{eq:power_est}
\end{equation}
where $N$ is the number of observed signal samples. {By assuming interference-free environment and equal-variance} noise at each receiver channel, the weights are only required to shift the phases of the received signals. Hence, they will have the same amplitude and can be written as
\begin{equation}
    \omega_i = 1/\sqrt{M} e^{j\theta_i}.
    \label{eq:omegatheta}
\end{equation}
Consequently, the derivative of the average power with respect to {each phase would be derived as}
\begin{equation}
    \frac{\partial {\hat{P}}}{\partial \theta_i} = \frac{1}{N} \sum_{n=0}^{N-1} \left(\frac{\partial y[n]}{\partial \theta_i}{y^*[n]} +\frac{\partial{y^*[n]}}{\partial \theta_i}y[n]\right) .
\end{equation}
Substituting for $y[n]$ from (\ref{eq:yn}) and for $\omega_i$ from (\ref{eq:omegatheta}) yields 
\begin{equation}
    \frac{\partial {\hat{P}}}{\partial \theta_i} = \frac{1}{N} \sum_{n=0}^{N-1} \left(j{r_i}[n]e^{j\theta_i}{y^*[n]} - j{r_i^*}[n]e^{-j\theta_i}y[n]\right).
    \label{eq:derivative}
\end{equation}
\subsection{Weight Update Process}
Starting from a random initial {weight vector} {$\boldsymbol{\omega}^0$}, the algorithm will update the weights at each data frame of size $N'$, and all frame samples can be used for the gradient estimation ($N'=N$) or $N$ samples can be used such that $N < N'$. Having the previous {weight vector} {$\boldsymbol{\omega}^{k-1}$},  the weights will then be updated at the {$k^{th}$} frame as 

\begin{equation}
    \omega_i^{k} = \frac{1}{\sqrt{M}} \exp\left\{{j\left(\theta_i[k-1]+\mu\frac{\partial P[k]}{\partial \theta_i[k-1]}\right)}\right\},
    \label{eq:weight_update}
\end{equation}
where $\mu$ is the step size. Expressing the weights in this form ensures that the {weight vector} has a unit norm and no normalization is required. {The brief description of the gradient-based beamforming algorithm} is given in Algorithm \ref{alg:BF}.
\begin{algorithm}
\caption{Gradient-Based Beamforming Algorithm}
\begin{algorithmic}[1]
\State initialize $\boldsymbol{\omega}^{0}$, define $\mu $
\While {receiving data}
\State calculate $y[1:N]=\boldsymbol{\omega}^H{\mathbf{r}}[1:N]$
\State estimate $\hat{P}$ using eq. (\ref{eq:power_est})
\State {evaluate the derivative ${\partial \hat{P}}/{\partial \theta_i}$} as given in eq. (\ref{eq:derivative})
\State update the {weights} according to eq. (\ref{eq:weight_update})
\State generate output $y[1:N']=\boldsymbol{\omega}^H{\mathbf{r}}[1:N']$
\EndWhile
\end{algorithmic}
\label{alg:BF}
\end{algorithm}
At the beginning, the weights can be initialized to take any random values. However, in simulations, the initial value for each weight {has been selected} as $\omega_i = 1/\sqrt{M}$. This initialization has shown well adaptiveness performance, thus it is preferred over random initialization to avoid possible convergence problems. 
In the following section, a detailed analysis of the performance of the proposed algorithm, its convergence, and adaptiveness is conducted.
\section{Numerical \& Experimental Results}
\begin{figure}
\centering
\subfloat[$8$-element ULA with QPSK]{%
\includegraphics[clip,width=0.99\columnwidth]{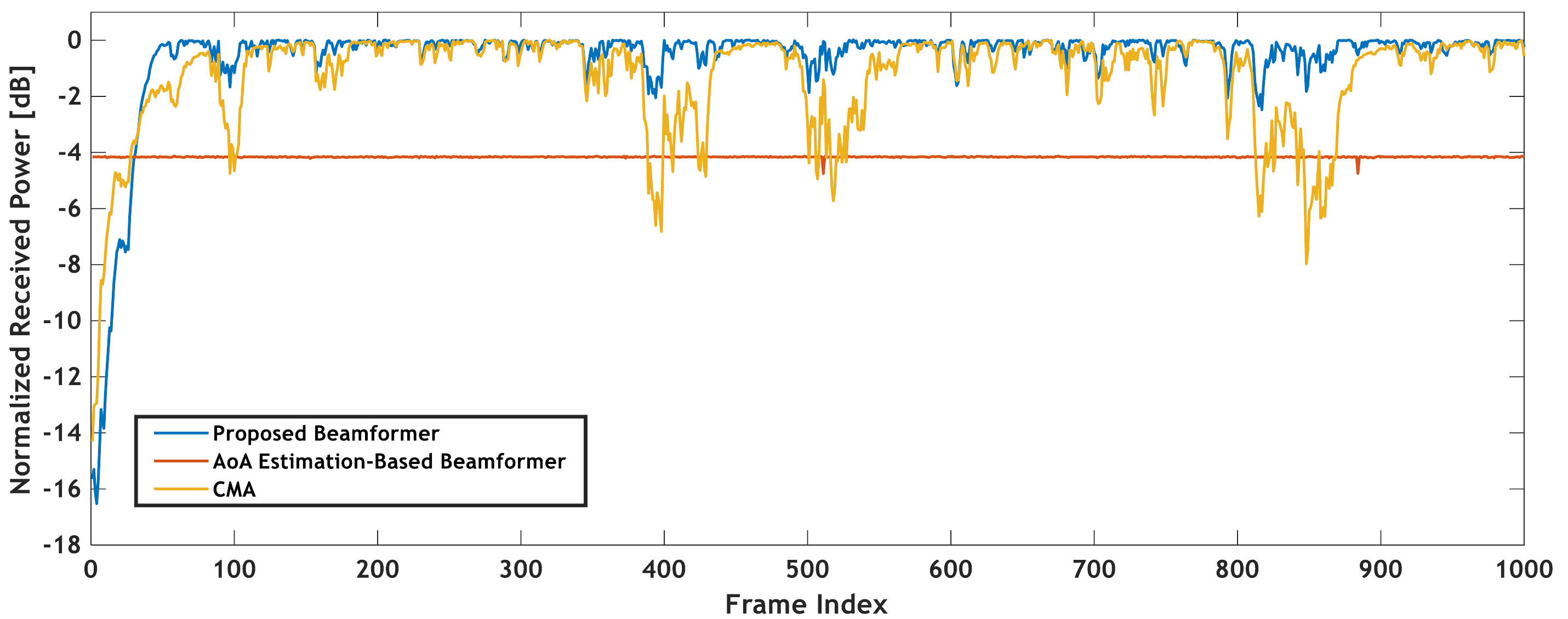}%
} \\
\subfloat[$8$-element UCA with QPSK]{%
\includegraphics[clip,width=0.99\columnwidth]{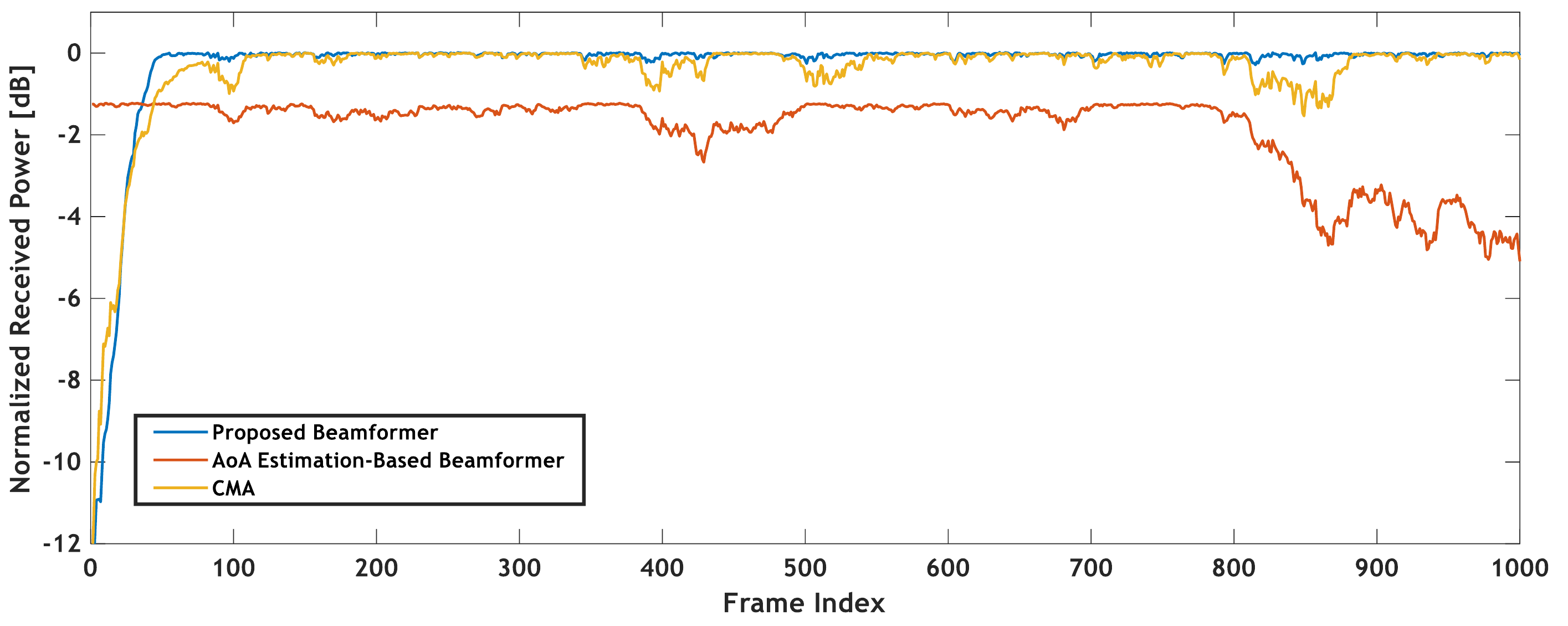}%
} \\ 
\caption{The beamformer output power versus received data frames.}
\label{fig:performance1}
\vspace{-2pt}
\end{figure}

This section exhibits the outcomes of the Monte Carlo simulations and the anechoic chamber measurements corresponding to both ULA and UCA configurations. Within our comparative study, the carrier frequency is selected as {$f_c=2$ GHz}, the inter-element distances are set equal to the half-wavelength corresponding to $f_c$, and the modulation scheme is selected as QPSK. The swept parameters in the simulations are SNR, AoA, and $N$. While evaluating the convergence of the beamformer and its variation due to the number of considered samples, the value of the AoA has been fixed, and different ($N$, SNR) combinations have been simulated for both array configurations. Here, each simulated baseband symbol has been oversampled by a factor of $8$ with the purpose of transmit filtering application. As a result of the beamformer's convergence examinations, choosing $N$ equal to samples per symbol (SPS) that is identical to the oversampling factor $8$ (one symbol) has been shown to be adequate for the proposed algorithm. On average, it takes up to $25$ iterations to reach the optimum weight vector. On the other hand, the variations in SNR changes have been shown to induce negligible effects on the overall performance. The time-varying AoA related to the signal source has been then modeled as a random walk process. Fig. \ref{fig:performance1} depicts the normalized average power of the proposed blind beamformer together with the outcomes of the conventional MUSIC approach and the prominent state-of-the art beamformer approach (i.e., CMA [4]). For the transmitted QPSK symbols, the advantages achieved by the proposed approach and the CMA in comparison to the fragile MUSIC algorithm have been exhibited due to successive data frame indices for both ULA and UCA configurations with $8$ elements. Here, note that, the received power has been normalized to the optimum solution given in (\ref{eq:optimum_solution}). Unless otherwise specified, the first $8$ symbols of each frames have been utilized for the beamforming algorithms. 
For the proposed model, after the transient period introduced during the convergence of the weight vector, the steady-state weights are obtained in the form of 
\begin{equation}
    \boldsymbol{\hat{\omega}} = \mathbf{a}_{I}(\phi) e^{j\gamma},
\end{equation} 
where $\gamma$ is an arbitrary phase shift. Consequently, this solution is shown to achieve maximum power at the beamformer's output. As seen in the Fig. \ref{fig:performance1}, the proposed algorithm outperforms the other techniques in terms of convergence to the optimum solution and convergence speed. While CMA at some instances struggles to modify the weights adaptively in response to changing AoA, the suggested model can monitor the changes in AoA with some minor dips in the ULA case.

\begin{figure}[b]
\centering
\includegraphics[width=0.99\columnwidth]{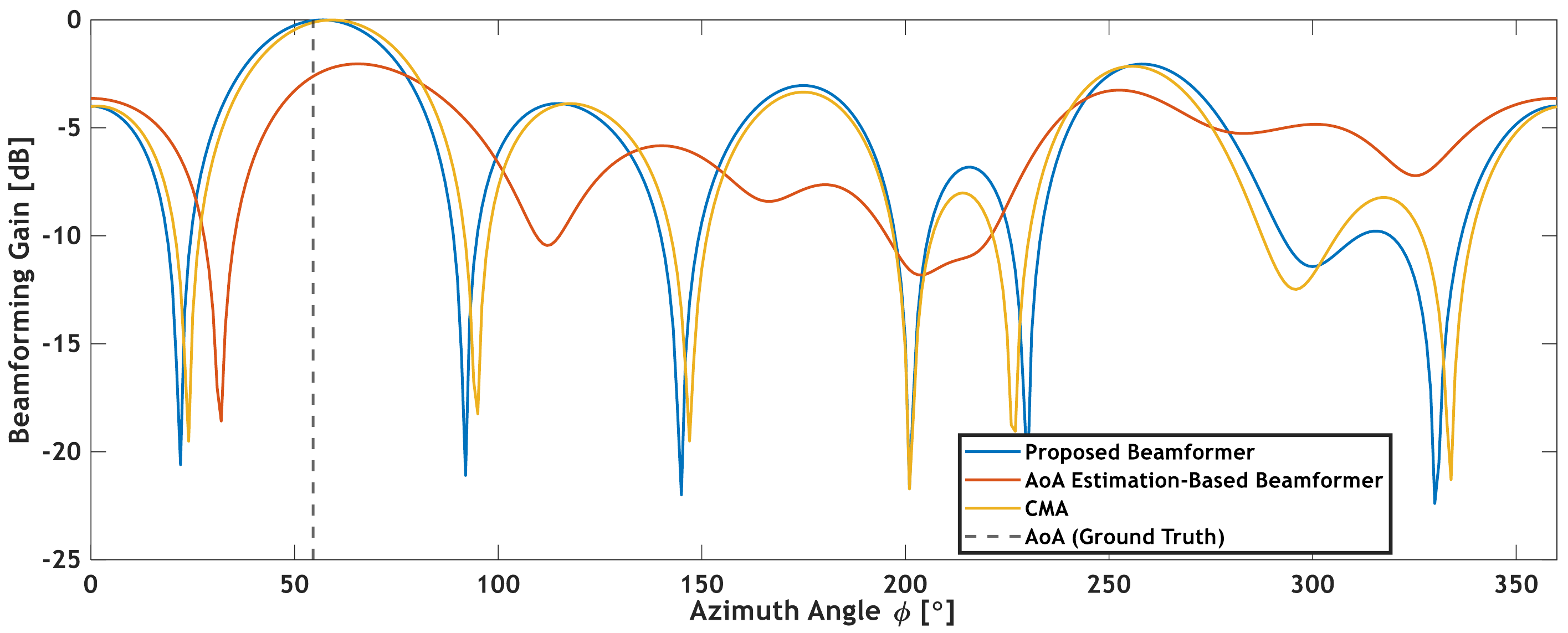}
\caption{$8$-element UCA array pattern with different beamformers.}
\label{fig:patterns}
\end{figure}

The formed beams for different approaches are shown in Fig. \ref{fig:patterns}. As seen, the proposed algorithm provides generating the highest peak in the AoA, and CMA has been shown to construct a very similar pattern. Here, the beamformer weights are compensated by the calibration coefficients to result in a real azimuth pattern.

\begin{figure}[t]
\centering
\includegraphics[width=0.99\columnwidth]{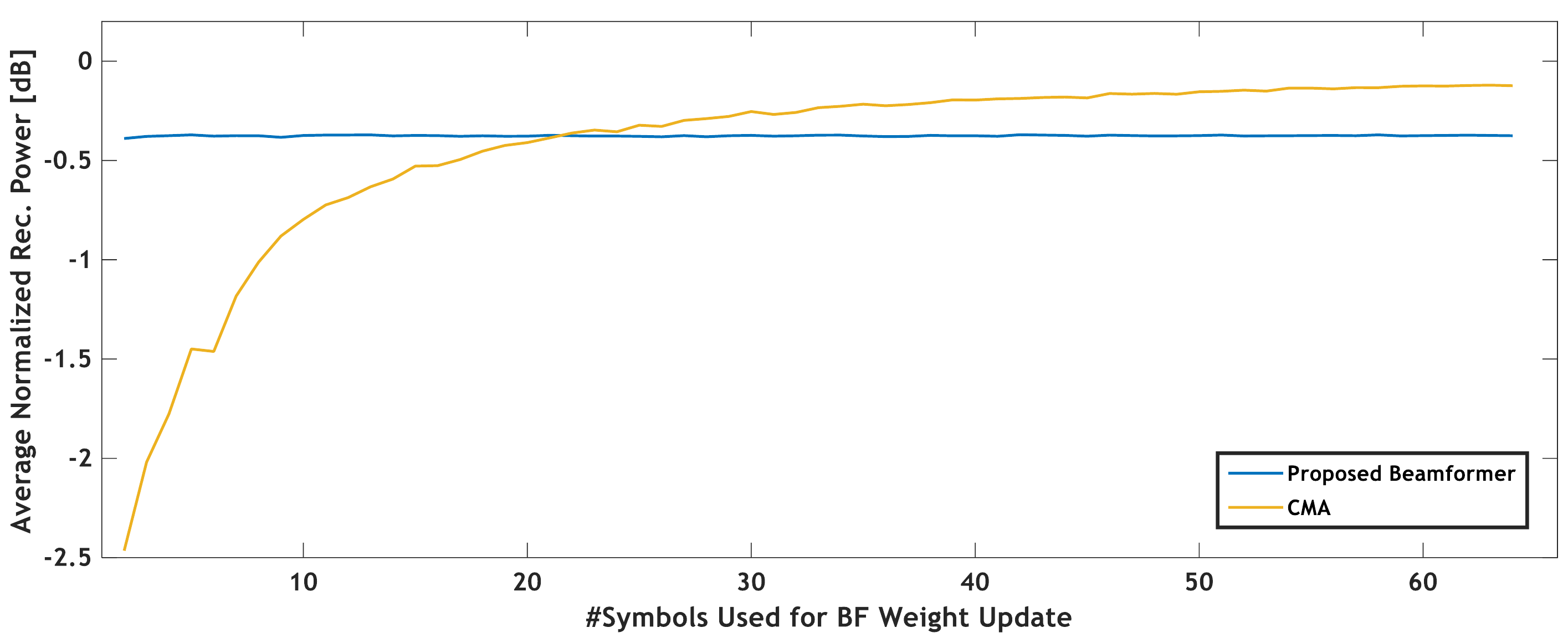}
\caption{Variation of average normalized output powers with respect to the number of symbols used for the beamforming process.}
\label{fig:avg_power}
\end{figure}

Additionally, Fig. \ref{fig:avg_power} shows the effect of varying the number of symbols for the weight update process on the performance. The vertical axis denotes the average power of beamformers' outputs for all frames normalized with respect to the optimum solution. As clearly seen, the proposed model exhibits considerable robustness even with lower number of observation samples. Here, note that, the deviation with respect to the optimum solution is caused due to the consideration of the first frames before convergence. For more than $20$ symbols, CMA converges faster and produces better solution compared to the proposed model. However, it is important to remember the CMA treats symbols iteratively, and the number of iterations is equal to the number of symbols, while the proposed model performs a single iteration at each frame using all the symbols at once. This clearly emphasizes the advantages of the proposed blind beamformer as a lower complexity but efficient solution when compared to the other state-of-art alternatives.

\begin{figure}
\centering
\includegraphics[width=0.95\columnwidth]{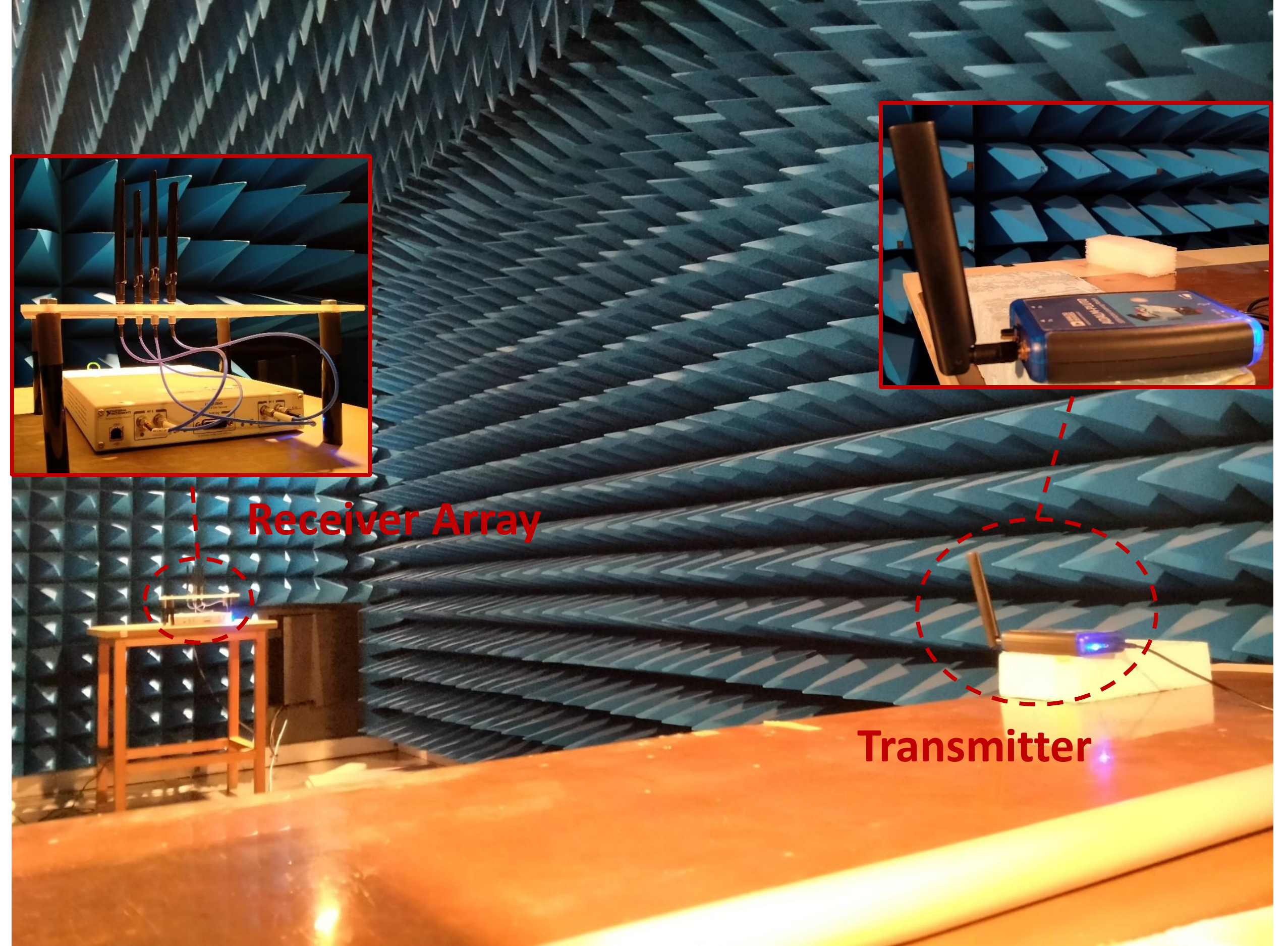}
\caption{The experimental setup used to verify the proposed blind beamformer.}
\label{fig:exp_setup}
\end{figure}

The performance of the proposed beamformer was evaluated in an anechoic chamber using the experimental setup shown in Fig. \ref{fig:exp_setup}. 
\begin{figure}
\centering
\includegraphics[width=0.99\columnwidth]{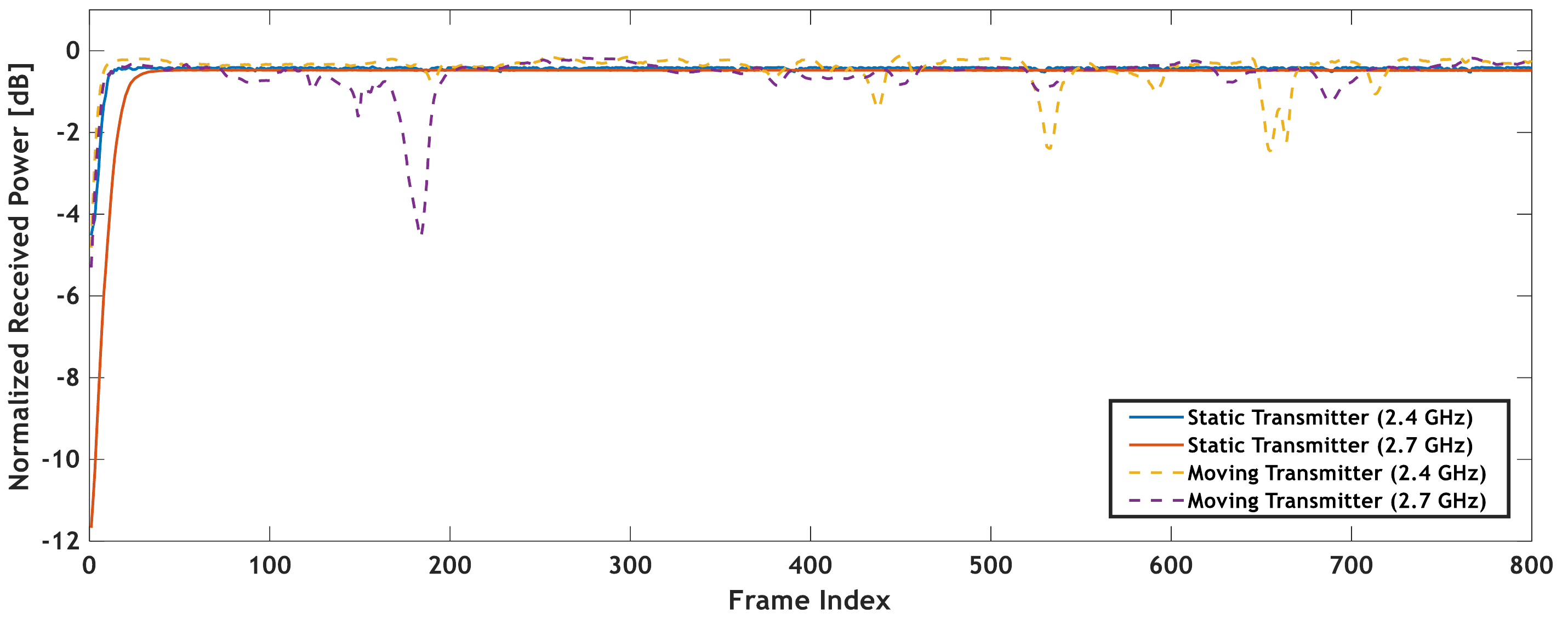}
\caption{The measured beamformer's ouput.}
\label{fig:exp_res}
\end{figure}
A receiver array composed of $4$ identical omnidirecitonal antennas \cite{antenna} and a transmitter equipped with a single antenna were utilized. The signals were received using USRP-2945 SDR with $4$ coherent channels and the proposed beamformer was implemented via a LabView environment. The measurements have been taken in $2.4$ GHz and $2.7$ GHz frequencies. For the $2.7$ GHz case, the UCA shown in the figure with interelement distance equal to $4.5 cm$ was utilized. For the $2.4$ GHz case, a ULA with interelement distance of $5$ cm was utilized. A QPSK-modulated signal with $4$ MHz bandwidth has been transmitted, the measurements have been performed for both the static transmitter case together with the moving transmitter case. The measurement results have been depicted in Fig. \ref{fig:exp_res}.
The proposed model has been shown to converge in a fast manner and keep the performance in the static case. Whereas, for the case of a moving transmitter, performance degradations have been introduced despite the beamformer being able to reconverge in short time periods ($\sim20$ ms).

\section{Conclusion}
In this paper, a low-complexity blind receiver beamforming approach has been introduced. This approach has been shown to converge to the optimum solution numerically, and verified by anechoic chamber measurements. Without prior information on the array's geometry, and in the presence of unknown mutual coupling and miscalibration errors, the beamformer was able to adaptively converge to the maximum power level giving weights. The proposed scheme has been tested for both ULA- and UCA-type antenna arrays. Since the weights found by the beamformer are very close to the optimum weights, as a future work, designing an algorithm that estimates the AoA from the weight vector under the specified imperfections would be beneficial. At the system level, estimating the AoA and receiver nonidealities could be significant for transmitter beamforming, which depends on receiver's feedback. 




%
\bibliographystyle{IEEEtran}

\bibliography{refs}

\end{document}